\documentclass[twocolumn,showpacs,preprintnumbers,amsmath,amssymb,superscriptaddress,nofootinbib,english]{revtex4-1}

\usepackage{graphicx}
\usepackage{dcolumn}
\usepackage{bm}
\usepackage{epsfig}
\usepackage{graphicx}
\usepackage{hyperref}
\usepackage[usenames]{color}
\usepackage{url}

\hypersetup{
    colorlinks=true,
    linkcolor=green,
    citecolor=blue,
}

\newcommand{\remove}[1]{}

\def\ie{{\frenchspacing\it i.e.}}

\def\be{\begin{equation}}
\def\ee{\end{equation}}
\def\ba{\begin{eqnarray}}
\def\ea{\end{eqnarray}}

\begin{document}

\title{Nonlinear Structure Formation with the  Environmentally Dependent Dilaton}

\author{Philippe~Brax}
\email[Email address: ]{philippe.brax@cea.fr}
\affiliation{Institut de Physique Theorique, CEA, IPhT, CNRS, URA 2306, F-91191Gif/Yvette Cedex, France}

\author{Carsten~van~de~Bruck}
\email[Email address: ]{c.vandebruck@sheffield.ac.uk}
\affiliation{Department of Applied Mathematics, University of Sheffield, Hounsfield Road, Sheffield S3 7RH, UK}

\author{Anne-Christine.~Davis}
\email[Email address: ]{a.c.davis@damtp.cam.ac.uk}
\affiliation{DAMTP, Centre for Mathematical Sciences, University of Cambridge, Wilberforce Road, Cambridge CB3 0WA, UK}

\author{Baojiu~Li}
\email[Email address: ]{b.li@damtp.cam.ac.uk}
\affiliation{DAMTP, Centre for Mathematical Sciences, University of Cambridge, Wilberforce Road, Cambridge CB3 0WA, UK}
\affiliation{Kavli Institute for Cosmology Cambridge, Madingley Road, Cambridge CB3 0HA, UK}

\author{Douglas~J.~Shaw}
\email[Email address: ]{d.j.shaw@damtp.cam.ac.uk}
\affiliation{DAMTP, Centre for Mathematical Sciences, University of Cambridge, Wilberforce Road, Cambridge CB3 0WA, UK}

\date{\today}

\begin{abstract}

We have studied the nonlinear structure formation of the environmentally dependent dilaton model using $N$-body simulations. We find that the mechanism of suppressing the scalar fifth force in high-density regions works very well. Within the parameter space allowed by the solar system tests, the dilaton model predicts small deviations of the  matter power spectrum and the mass function from their $\Lambda$CDM counterparts. The importance of taking full account of the nonlinearity of the model is also emphasized.

\end{abstract}

\maketitle

\section{Introduction}

Modifying gravity on large scales is one of the plausible ways of explaining the recent acceleration of the expansion of the universe. So far, the construction of valid models of modified gravity has been fraught with difficulties. The most serious one is already present in the original Pauli-Fierz formulation of massive gravity \cite{rubakov} and involves the existence of a ghost in curved backgrounds\footnote{For a new formulation of massive gravity in the so called decoupling limit see \cite{Tolley}}. This phenomenon seems to be generic as suggested by Ostrograski's theorem\cite{woodard} which states that  higher derivative theories have a Hamiltonian which is unbounded from below. Higher dimensional versions of modified gravity such as the DGP model \cite{DGP} also suffer from the presence of a ghost in their spectrum at low energy. This problem is nicely avoided in $f(R)$ models \cite{fR} which turn out to be equivalent to a particular type of scalar-tensor theories \cite{scalar_tensor}. In these models, the compatibility with solar system and laboratory tests of gravity is not straightforward and can only be achieved thanks to the so-called chameleon mechanism \cite{Khoury:2004a, Khoury:2004, Mota:2006, Mota:2007, Li:2007, Brax:2008}. Indeed, the existence of a nearly massless scalar field on cosmological scales could jeopardize gravity locally. This issue is common to all known models of dark energy coupled to matter \cite{Carroll}. In a large class of dark energy models involving a linear coupling to matter and a non-linear potential, the chameleon mechanism, whereby the scalar field mass becomes dependent on the ambient environment, would be sufficient to hide away the dark energy field locally. Similarly, in models of gravity such as the DGP or Galileon \cite{Chow:2009, DeFelice:2010} theories for which a shift symmetry only allow for non-linearities in the scalar field kinetic terms, the Vainshtein mechanism \cite{Vainshtein:1972} can be at play and prevent the existence of a fifth force locally. In this paper, we will focus on a different type of models involving a scalar field. These models are inspired from the string dilaton in the strong coupling regime~\cite{dilaton1,dilaton2,dilaton3}. Their gravitational validity relies on an environmentally dependent form of the Damour-Polyakov mechanism \cite{Damour:1994} whereby the coupling to matter is driven to vanish cosmologically. Here, the coupling to matter is negligible in dense regions and in the vicinity of dense bodies. This prevents the existence of a fifth force in galaxies. Constraints on the parameter space of these models springing from local tests of gravity have already been obtained in \cite{Brax:2010}. Here we will study the cosmology of these models in the non-linear regime when structures form (see \cite{Brax:2011} for an analysis of the bispectrum of matter distribution in this model). This requires large computer simulations. As a result, we have access to non-linear properties of the dilaton models such as the non-linear power spectrum or the number of dark matter halos for a given mass. Moreover we will be able to probe how much the local tests of gravity constrain large scale structure formation and deviations from general relativity. We find that the dilaton models differ from GR at most at the level of a few percent once the local (i.e. solar--system) constraints have been imposed. Although possibly detectable in principle, observing such small deviations will be challenging in the near future.

The arrangement of this paper is as follows: in Sect.~\ref{sect:model} we briefly review the dilaton model under study and derive the field equations in the Newtonian limit, which are relevant for the study of structure formation on sub-horizon scales and at late times. In Sect.~\ref{sect:Nbody} we describe the algorithm and code of our $N$-body simulations, and perform relevant tests of the code. More technical details are given in the appendices. The numerical results and their analysis are summarised in Sect.~\ref{sect:results} and finally we conclude in Sect.~\ref{sect:con}. The metric convention is $(-,+,+,+)$ and we use $c=1$ unless stated otherwise.

\section{The Environmentally Dependent Dilaton}

\label{sect:model}

In this section we very briefly summarise the essential ingredients of the environmentally dependent dilaton model, which will be used for the simulations and discussions below. For more details about the model the reader is referred to \cite{Brax:2010}.

\subsection{The Model}

The dilaton model is fully specified by the following Einstein-Hilbert action in the Einstein frame:
\ba\label{eq:action}
\mathcal{S} &=& \int\sqrt{-g}d^4x\left[\frac{R}{2\kappa^2}-\frac{k^2(\varphi)}{\kappa^2}\nabla^a\varphi\nabla_a\varphi-V(\varphi)\right]\nonumber\\
&&+\mathcal{S}_{\rm{m}}\left(\Psi_i,A^2(\varphi)g_{ab};\varphi\right),
\ea
in which $g$ is the determinant of the metric $g_{ab}$, $\kappa^2\equiv8\pi G$ with $G$ the gravitational constant, $\varphi$ is the dilaton field,  and $V(\varphi)$ its potential, which is derived from string theory in the strong coupling limit. In the matter action $\mathcal{S}_{\rm{m}}$, $\Psi_i$ collectively represents the matter fields and $A^2(\varphi)g_{ab}$ is the metric governing the geodesics of matter particles. In the Einstein frame, the particles feel an extra, or fifth, force whose strength is determined by the coupling function $\beta(\varphi)\equiv\left[\ln A(\varphi)\right]_{,\varphi}$ where a comma denotes partial differentiation. The function $k(\varphi)$ is given by
\ba\label{eq:k_def}
k(\varphi) &\approx& \lambda^{-1}\sqrt{1+3\lambda^2\beta^2\varphi}
\ea
where $\lambda$ is a constant. Throughout this paper Latin indices $a,b,c,\dots$ run over $0,1,2,3$ and Greek indices $\alpha,\beta,\dots$ run over 1,2,3.

Varying the action with respect to the metric $g_{ab}$, we obtain the total energy-momentum tensor of the model,
\ba\label{eq:T_ab}
\kappa^2T_{ab} &=& \kappa^2A(\varphi) T^{\rm{m}}_{ab} - \kappa^2g_{ab}V(\varphi)\nonumber\\
&&+ k^2(\varphi)\left[2\nabla_a\varphi\nabla_b\varphi-g_{ab}\nabla^c\varphi\nabla_c\varphi\right]
\ea
where $T^{\rm{m}}_{ab}$ is the energy-momentum tensor for fluid matter, \ie, baryons, radiation and cold dark matter (CDM).  Note that there is a factor $A(\varphi)$ in front of $T^{\rm{m}}_{ab}$.  $T_{\rm{m}}^{ab}$ is not, in general, conserved but instead:
\ba 
\nabla_{a} T_{\rm m}^{ab} = \frac{A_{,\varphi}(\varphi)}{A(\varphi)} \left[ T_{\rm m}\nabla^{b}\phi - T_{\rm m}^{ab} \nabla_{a}\phi\right]. \label{eq:econs}
\ea
For pressureless dust, where $T_{\rm m}^{ab} = \rho_{\rm m} u^{a} u^{b}$, $u_{a}u^{a}=-1$, Eq. (\ref{eq:econs}) implies that the usual continuity equation holds, $\nabla_{a}(\rho_{\rm m}u^{a}) = 0$, and hence $\rho_{\rm m}$ is conserved.  In a Robertson--Walker spacetime, this means that the usual conservation equation for matter still holds:
\ba
\dot{\bar{\rho}}_{m} + 3H\bar{\rho}_m &=& 0,
\ea
in which $\dot{}\equiv d/dt$, subscript $_{m}$ denotes matter, $H=\dot{a}/{a}$ is the background expansion rate with $a$ the scale factor, and an overbar stands for the background value of a physical quantity. The gravitational field equation, or Einstein's equation, is given as usual:
\ba\label{eq:Einstein}
G_{ab} \equiv R_{ab}-\frac{1}{2}g_{ab}R = \kappa^2T_{ab},
\ea
where $G_{ab}, R_{ab}$ and $R$ are respectively the Einstein tensor, Ricci tensor and Ricci scalar.

Varying the action with respect to the scalar field $\varphi$, we obtain its equation of motion:
\ba\label{eq:sfeom}
&&\nabla^a\left[k(\varphi)\nabla_a\varphi\right]\nonumber\\
&=& \frac{4\pi G}{k(\varphi)}\left[-V(\varphi)-\beta(\varphi)\left(A(\varphi) T^{\rm{m}}-4V(\varphi)\right)\right]
\ea
where $T^{\rm{m}}$ is the trace of $T^{\rm{m}}_{ab}$.
The energy-momentum tensor of an individual particle with mass $m_0$ at position $\bf{r}_0$ is given by
\ba
T^{ab}_m(\bf{r}) &=& \frac{m_0}{\sqrt{-g}}\delta\left(\bf{r}-\bf{r}_0\right)\dot{r}^a_0\dot{r}^b_0,
\ea
where $\bf{r}$ is the general spatial coordinate. Using the Bianchi identity we get
\ba\label{eq:geodesic}
\ddot{r}^a_0 + \Gamma^a_{bc}\dot{r}^b_0\dot{r}^c_0 &=& -\beta(\varphi)\nabla^a\varphi -\beta(\varphi)\dot{\varphi}\dot{r}^a_0,
\ea
in which $\Gamma^a_{bc}$ is the Levi-Civita connection. Clearly, if $\beta=0$ then this reduces to the geodesic equation in general relativity, as expected.

Eqs.~(\ref{eq:Einstein}, \ref{eq:T_ab}, \ref{eq:sfeom}, \ref{eq:geodesic}) contain all the physics for the analysis below, though to implement them in $N$-body simulations we still have to simplify them using appropriate approxi\-mations. These will be carried out below.

In this paper, we focus on the particular model of \cite{Brax:2010}, which is
motivated from string theory, specified by
\ba\label{eq:model}
A(\varphi) &=& 1+\frac{1}{2}A_2\left(\varphi-\varphi_0\right)^2,\\
\beta(\varphi) &=& A_2\left(\varphi-\varphi_0\right),\\
k^2(\varphi) &=& 3A_2^2\left(\varphi-\varphi_0\right)^2+\lambda^{-2},\\
V(\varphi) &=& A^4(\varphi)V_0e^{-\varphi}
\ea
where $A_2\gg1$ is a parameter and $\varphi_0$ is the current background value of $\varphi$. $V_0$ is another  parameter of mass dimension 4. Because the potential is exponential, we are free to shift the value of $\varphi$ so that $\varphi_0=0$. Clearly $V_c=V_0e^{-\varphi_0}=V_0$ should be chosen carefully so that it can play the role of dark energy today. Both the parameters $A_2$ and $\lambda$ are crucially constrained by local tests. In the numerical simulations, we will choose values of the parameters which are on the verge of the allowed parameter space in order to enhance the possible effects on large scales.

\subsection{Nonrelativistic Limit}

Eqs.~(\ref{eq:Einstein}, \ref{eq:T_ab}, \ref{eq:sfeom}, \ref{eq:geodesic}) are general relativistic equations.  To implement them into $N$-body simulations for large scale structure formation, it suffices to work in the non-relativistic limits, since the simulations only probe weak-gravity regime and small volumes compared with the cosmos.

We write the perturbed metric in the conformal Newtonian gauge as
\ba\label{eq:metric}
ds^2 &=& -a^2(1+2\phi)d\tau^2 + a^2(1-2\psi)\gamma_{\mu\nu}dx^\mu dx^\nu
\ea
where $\tau, x^\mu$ are respectively the conformal time and comoving coordinate, $\gamma_{\mu\nu}$ is the metric of a 3-D Euclidean space, and $\phi, \psi$ respectively the Newtonian potential and the perturbation to the spatial curvature. For completeness, we list the expressions of the components of $G_{ab}$ in terms of the metric variables using our convention in Appendix~\ref{appen:express}.

Let us first look at the scalar field equation of motion Eq.~(\ref{eq:sfeom}). For this, we define $\xi$ such that $\nabla_a\xi = k(\varphi)\nabla_a\varphi$, and write $\nabla^a\left[k(\varphi)\nabla_a\varphi\right]$ to first order in the metric perturbation variables as
\ba
a^2\nabla^a\left[k(\varphi)\nabla_a\varphi\right] &\approx& -(1-2\phi)\xi'' + \nabla^2_{\bf{x}}\xi\nonumber\\
&&- \xi'\left[2\frac{a'}{a}(1-2\phi)-(\phi'+3\psi')\right]\nonumber
\ea
where $'\equiv d/d\tau$, and $\nabla_{\bf{x}}$ is the derivative with respect to the comoving coordinate $\bf{x}$. Substituting this expression into Eq.~(\ref{eq:sfeom}), and removing the background equation of motion
\ba
&&\left[k(\bar{\varphi})\bar{\varphi}'\right]' + 2\frac{a'}{a}k(\bar{\varphi})\bar{\varphi}'\nonumber\\
&=& \frac{4\pi Ga^2}{k(\bar{\varphi})}\left[V(\bar{\varphi})-\beta(\bar{\varphi})\left(A(\bar{\varphi})\bar{\rho}_{\rm{m}}+4V(\bar{\varphi})\right)\right],
\ea
we obtain the perturbation part of this equation:
\ba\label{eq:sfeom_nr}
&&\nabla_{\bf{x}}\cdot\left[k(\varphi)\nabla_{\bf{x}}\varphi\right]\nonumber\\
&\approx& \frac{4\pi Ga^2}{k(\varphi)}\left\{\beta(\varphi)\left[A(\varphi)\rho_{\rm{m}}+4V(\varphi)\right]-V(\varphi)\right\}\nonumber\\
&&-\frac{4\pi Ga^2}{k(\bar{\varphi})}\left\{\beta(\bar{\varphi})\left[A(\bar{\varphi})\bar{\rho}_{\rm{m}}+4V(\bar{\varphi})\right]-V(\bar{\varphi})\right\}.
\ea
Note that in the above derivation we have dropped terms such as $\phi', \psi'$ and $\frac{a'}{a}\phi$, since we are working in the quasi-static limit in which the time derivative of a quantity is much smaller than its spatial gradient, \ie, $|\nabla_{\bf{x}}\phi|\gg|\phi'|$.

Using the expressions given in Appendix~\ref{appen:express}, we can write the $00$-component of the Ricci scalar as
\ba
a^2R^0_{\ 0} &\approx& -\nabla^2_{\bf{x}}\phi + 3\left[\frac{a''}{a}-\left(\frac{a'}{a}\right)^2\right](1-2\phi)\nonumber\\
&&- 3\psi'' - 3(1-2\phi)\frac{a'}{a}\left(\phi'+\psi'\right)\nonumber
\ea
again up to first order in the perturbed metric variables. Similarly,
\ba
8\pi GT &\approx& -8\pi G\left[A(\varphi)\rho_{\rm{m}}+4V(\varphi)\right]\nonumber\\
&&+ k^2(\varphi)\frac{2}{a^2}(1-2\phi)\varphi'^2\nonumber
\ea
where $T$ is the trace of the total energy-momentum tensor.  Then the $00$-component of the Einstein equation
\ba
R_{ab} &=& 8\pi G\left(T_{ab}-\frac{1}{2}g_{ab}T\right),
\ea
with the background part, \ie, the Raychaudhuri equation,
\ba
3\left[\frac{a''}{a}-\left(\frac{a'}{a}\right)^2\right] &=& -4\pi GA(\bar{\varphi})\bar{\rho}_{\rm{m}}a^2\nonumber\\
&&- 2k^2(\bar{\varphi})\bar{\varphi}'^2 + 8\pi GV(\bar{\varphi})a^2
\ea
removed, can be written as
\ba\label{eq:poisson_nr}
\nabla_{\bf{x}}^2\Phi &\approx& 4\pi G\left[A(\varphi)\rho_{\rm{m}}-A(\bar{\varphi})\bar{\rho}_{\rm{m}}\right]a^3\nonumber\\
&&-8\pi G\left[V(\varphi)-V(\bar{\varphi})\right]a^3,
\ea
where we have defined $\Phi\equiv a\phi$ for convenience.

Finally, for the equation of motion of matter particles, Eq.~(\ref{eq:geodesic}), using the relationship between physical coordinates $\bf{r}$ and comoving distance $\bf{x}$, we can rewrite it as
\ba
\ddot{\bf{x}} + 2\frac{\dot{a}}{a}\dot{\bf{x}} &=& -\frac{1}{a^3}\nabla_{\bf{x}}\Phi - \frac{1}{a^3}\nabla_{\bf{x}}(a\varphi) - \beta\dot{\varphi}\dot{\bf{x}}.
\ea
Defining the conjugate momentum to $\bf{x}$ as $\bf{p}=a^2\dot{\bf{x}}$, this equation could be decomposed as
\ba\label{eq:dxdt}
\frac{d\bf{x}}{dt} &=& \frac{\bf{p}}{a^2},\\
\label{eq:dpdt}\frac{d\bf{p}}{dt} &=& -\frac{1}{a}\nabla_{\bf{x}}\Phi - \frac{1}{a}\beta(\varphi)\nabla_{\bf{x}}(a\varphi) - \beta(\varphi)\dot{\varphi}\bf{p}.
\ea
Note that there are two components of the fifth force, as discussed in \cite{Li:2010nc}

Eqs.~(\ref{eq:sfeom_nr}, \ref{eq:poisson_nr}, \ref{eq:dxdt}, \ref{eq:dpdt}) are all that we need to put into the $N$-body simulation code to study  structure formation in the nonlinear regime. Before that we have to discretise these equations and write them using code units, so that they can be applied on a mesh with finite grid size. These lengthy expressions are given in Appendix~\ref{appen:discrete}, where we also discuss the subtleties in the numerical implementation.

\section{The $N$-body Simulations}

\label{sect:Nbody}

In this section we briefly describe the algorithm and model specifications of the $N$-body simulations we have performed. We also give results for the tests of the code, which show that the scalar-field solver works quite well.

\subsection{Outline of the Simulation Algorithm}

For our simulations we have used a modified version of the publicly-available $N$-body code {\tt MLAPM} \cite{Knebe:2001}. The modifications we have made follow the detailed prescription of Ref.~\cite{Li:2010nc}, and here we only give a brief description.

The {\tt MLAPM} code has two sets of meshes: the first includes a series of increasingly refined regular meshes covering the whole cubic simulation box, with respectively $4, 8, 16, \cdots, N_d$ cells on each side, where $N_d$ is the size of the domain grid, which is the most refined of these regular meshes. This set of meshes are needed to solve the Poisson equation using multigrid method or fast Fourier transform (for the latter only the domain grid is necessary). When the particle density in a cell exceeds a predefined threshold, the cell is further refined into eight equally sized cubic cells; the refinement is done on a cell-by-cell basis and the resulting refinement could have arbitrary shape which matches the true equal-density contours of the matter distribution. This second set of meshes are used to solve the Poisson equation using the linear Gauss-Seidel relaxation scheme.

The dilaton field is the most important ingredient in the model studied here, and we have to solve it to obtain detailed information about the fifth force. In our $N$-body code, we have added a new scalar field solver which is based on Eqs.~(\ref{eq:diffop}, \ref{eq:diffop1}, \ref{eq:GS}, \ref{eq:diffop2}). It uses a nonlinear Gauss-Seidel scheme for the relaxation iteration and the same criterion for convergence as the default Poisson solver in {\tt MLAPM}. But it uses V-cycle \cite{Press:1988} instead of the self-adaptive scheme in arranging the Gauss-Seidel iterations.

The value of $u$ (see definition in Appendix~\ref{appen:discrete}) solved in this way is then used to calculate the total energy density including that of the scalar field, and this completes the computation of the source term to the modified Poisson equation. The latter is then solved using fast Fourier transform on the domain grid and Gauss-Seidel relaxation on refinements, according to Eq.~(\ref{eq:poisson_dis}).

With the gravitational potential $\Phi$ and the scalar field $u$ at hand, we can use Eq.~(\ref{eq:dpdt_dis}) to evaluate the total force on the particles and update their momenta/velocities. Then Eq.~(\ref{eq:dxdt_dis}) is used to advance the particles in space.

For more details about the implementation see \cite{Li:2010nc}.

\subsection{Simulation Details}

The physical parameters we use in the simulations are as follows: the present dark-energy fractional energy density $\Omega_{\Lambda}=0.743$ and $\Omega _{\rm{m}}=0.257$, $H_{0}=71.9$~km/s/Mpc, $n_{s}=0.963$ and $\sigma_{8}=0.769$. We use two sets of simulation box which have sizes of $32h^{-1}$~Mpc and $64h^{-1}$~Mpc respectively, in which $h=H_{0}/(100~\mathrm{km/s/Mpc})$. We simulate four models, with parameters $(A_2,\lambda)=(4\times10^6,2)$, $(4\times10^5,10)$, $(2\times10^5,100)$ and $(2\times10^6,30)$. These parameters are chosen so that they predict local fifth forces which are allowed by current experiments and observations\footnote{The values are taken from near the boundary of the allowed region in the parameter space in Fig.~1 of \cite{Brax:2010}. As a result we expect that they should give us the biggest effect on large-scale structure while satisfying constraint from local experiments.}. In all those simulations, the particle number is $256^{3}$, so that the mass
resolution is $1.114\times 10^{9}h^{-1}~M_{\bigodot }$ for the $64h^{-1}$~Mpc simulations and $1.393\times 10^{8}h^{-1}~M_{\bigodot }$ for the $32h^{-1}$~Mpc simulations. The domain grid is a $128\times
128\times128$ cubic and the finest refined grids have 16384 cells on each side, corresponding to a force resolution of about $12h^{-1}~$kpc and $6h^{-1}$~kpc respectively for the two sets of simulations. The force resolution determines the smallest scale on which the numerical results are reliable. We have also run a $\Lambda$CDM simulation with the same physical parameters.

Our simulations are purely $N$-body, which means that baryonic physics has not been included in the numerical code. We use the same initial conditions for the dilaton and the $\Lambda$CDM simulations, because before the initial redshift $z_i=49$ the fifth force is strongly suppressed so that the effect of the dilaton on the matter power spectrum is negligible.

\begin{figure}
\includegraphics[scale=0.5]{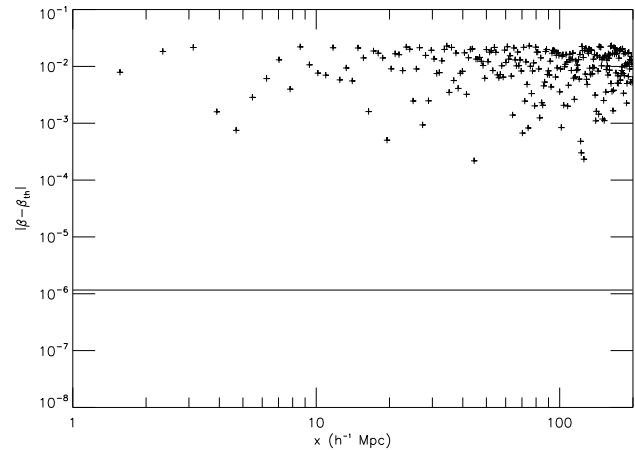}
\caption{A first test of the scalar field solver. For this test we use a simulation box of $256h^{-1}$~Mpc on each side, and set the density field to be homogeneous in the box. The exact value of $\beta$, $\beta_{\rm{th}}$, is known analytically. The differences between $\beta_{\rm{th}}$ the initial guess of $\beta$ in the grid cells along the $x$-axis are shown as symbols, while that between $\beta_{\rm{th}}$ and the $\beta$ after relaxation are shown as the continuous curve. Clearly the relaxation works accurately.}
\label{fig:test2}
\end{figure}

\begin{figure}
\includegraphics[scale=0.5]{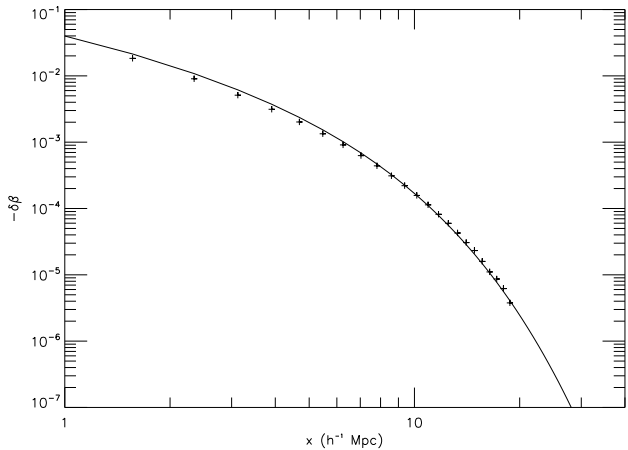}
\caption{A second test of the scalar field solver. For this test we use a simulation box of $128h^{-1}$~Mpc on each side, and set the density field to be the equivalent of having a point mass at $x=y=z=0$ and zero otherwise. Far from the point mass, the solution to $\beta$ can be approximately solved analytically (the continuous curve). The symbols show the results for $\beta$ from the numerical code. The two show good agreement in a wide range of $x$.}
\label{fig:test}
\end{figure}

\subsection{Code Tests}

Before displaying the numerical results from the $N$-body simulations, we show some evidence that our numerical procedure works correctly. As our modification to the default {\tt MLAPM} code is only in the scalar field part, we focus on tests of the scalar field solver and the fifth force only.

The scalar field solver uses the nonlinear Gauss-Seidel relaxation scheme to compute $\beta$, and an indicator that it works is to show that, given the initial guess of the solution that is very different from the true solution, the relaxation could produce the latter within a reasonable number of iterations. Consider a simulation box with homogeneous density, then the true solution to $\beta$, $\beta_{\rm{th}}$, could be calculated analytically. We therefore make an initial guess for $\beta$ which is randomly scattered around $\beta_{\rm{th}}$ and let the scalar solver try to recover $\beta_{\rm{th}}$. In Fig.~\ref{fig:test2} we have shown $|\beta-\beta_{\rm{th}}|$ before (symbols) and after (curve) the relaxation: as can be seen there, before the relaxation the difference between the initial guess $\beta$ and $\beta_{\rm{th}}$ is of order $0.01$, while after the relaxation it reduces to $10^{-6}$. Note that $10^{-6}$ corresponds to the error caused by using floating-point numbers, and as a result this shows that the scalar solver works accurately.

As a second test of the scalar field solver, consider having a point mass at the origin $x=y=z=0$ and the vacuum density otherwise. This could be achieved by filling the densities in the cells of the simulation grid according to \cite{Oyaizu:2008}
\ba
\rho_c &=& 10^{-4}N^3_d
\ea
for the cell with $i=j=k=0$, in which $N_d$ is the number of cells on each side of the domain grid, and $\rho_c=10^{-4}$ for all other cells.

Outside the particle it is the vacuum, in which the scalar field equation of motion can be approximately linearised as
\ba
\nabla^2_{\bf{x}}\delta\beta &\approx& 8\pi GV_0A_2 a^2\frac{3\bar{\beta}(1-2\bar{\beta})+2\lambda^{-2}}{\left(3\bar{\beta}^2+\lambda^{-2}\right)^2}\delta\beta,
\ea
where we remind the reader that $\beta = A_2\varphi$ (see eq.~\ref{eq:model} with $\varphi_0=0$), and $\delta\beta\equiv\beta-\bar{\beta}$. $\bar\beta$ is the background value of $\beta$, which can be analytically calculated as~\cite{Brax:2010}
\ba
\bar{\beta} &=& \frac{\Omega_\Lambda a^3}{\Omega_{\rm{m}}+4\Omega_\Lambda a^3}.
\ea
Using the code units (see Appendix~\ref{appen:discrete}), this can be written as
\ba
\nabla^2\delta\beta &\approx& m_{\rm eff}^2\delta\beta
\ea
with
\ba
m_{\rm eff}^2 &=& \frac{\left(BH_0\right)^2}{ac^2}3\Omega_\Lambda a^3A_2\frac{3\bar{\beta}(1-2\bar{\beta})+2\lambda^{-2}}{\left(3\bar{\beta}^2+\lambda^{-2}\right)^2}\delta\beta.
\ea
Here $B$ is the box size of the simulation box, and $c$ is the speed of light, which we have restored to make the dimension explicit. The analytic solution is thus
\ba\label{eq:delta_beta}
\delta\beta(r) &=& \frac{C}{r}e^{-m_{\rm eff}r}
\ea
where $C$ is some constant determined by the value of the point mass and $r$ the distance from the origin. Because $C$ is unknown, we fix its value by requiring that Eq.~(\ref{eq:delta_beta}) be equal to the numerical solution at $r=10h^{-1}$~Mpc.

The normalised analytical solution to $\beta$ is shown as the continuous curve in Fig.~\ref{fig:test}, while the numerical solutions are shown as symbols. We see that the two agree over a wide range of $r$ (note that $|\delta\beta|$ changes by several orders of magnitude). Note that when $r$ is small the agreement is not perfect, because linearisation does not work very well near the high density region; meanwhile, for very big $r$ the value of $|\delta\beta|$ drops below $\mathcal{O}(10^{-6})$ and numerical error due to using floating-point numbers becomes important.

In summary, Figs.~\ref{fig:test2} and \ref{fig:test} show that our scalar solver works well. Below, we also show that the fifth force agrees with analytic approximations in certain regimes.

\begin{figure*}
\includegraphics[scale=0.66]{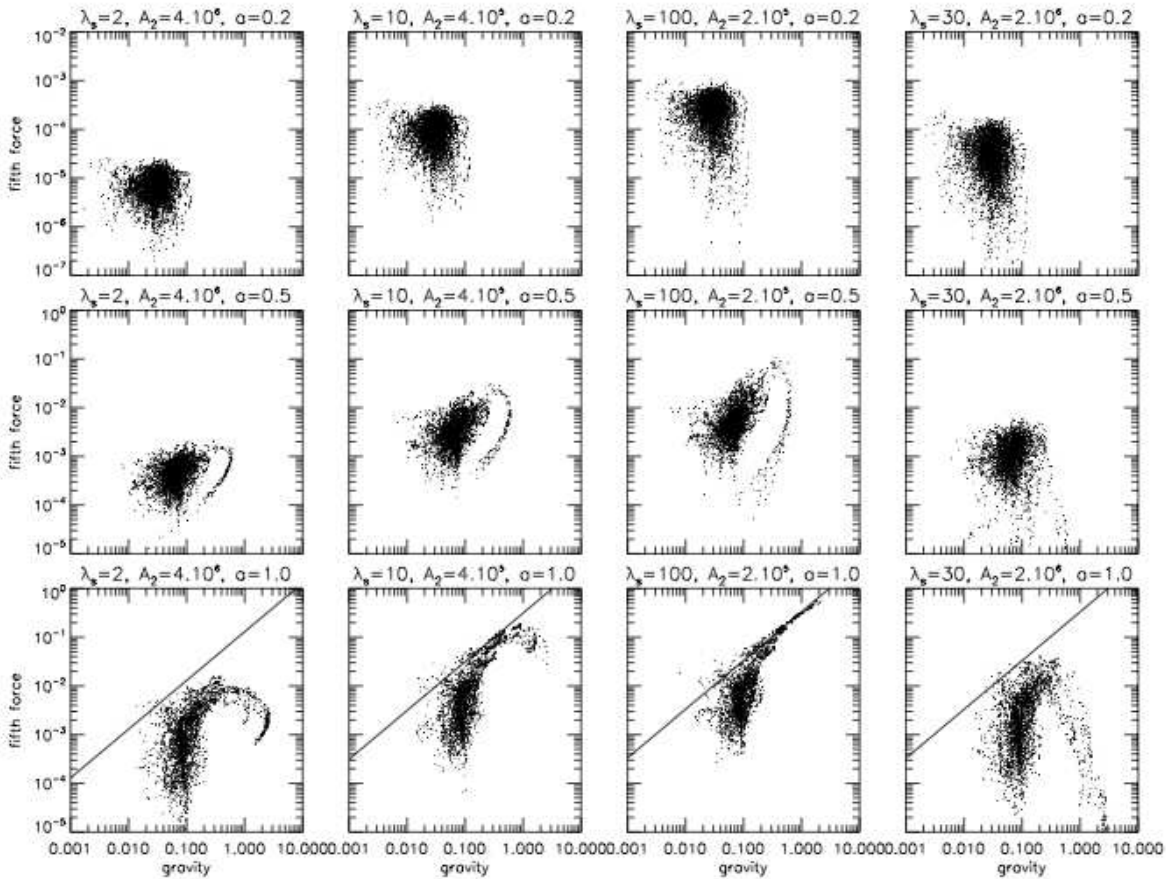}
\caption{The magnitude of the fifth force (the vertical axis) versus that of gravity (the horizontal axis), for the particles (black points) selected from a thin slice of the $32h^{-1}$~Mpc simulation box. We show this for the four models we have simulated and at three different output times, as given in the subtitle of each panel. Note that both forces are expressed using the internal unit (see Appendix~\ref{appen:discrete}), which is $H_0^2/B$ times the physical force unit.}
\label{fig:force}
\end{figure*}

\begin{figure*}
\includegraphics[scale=0.66]{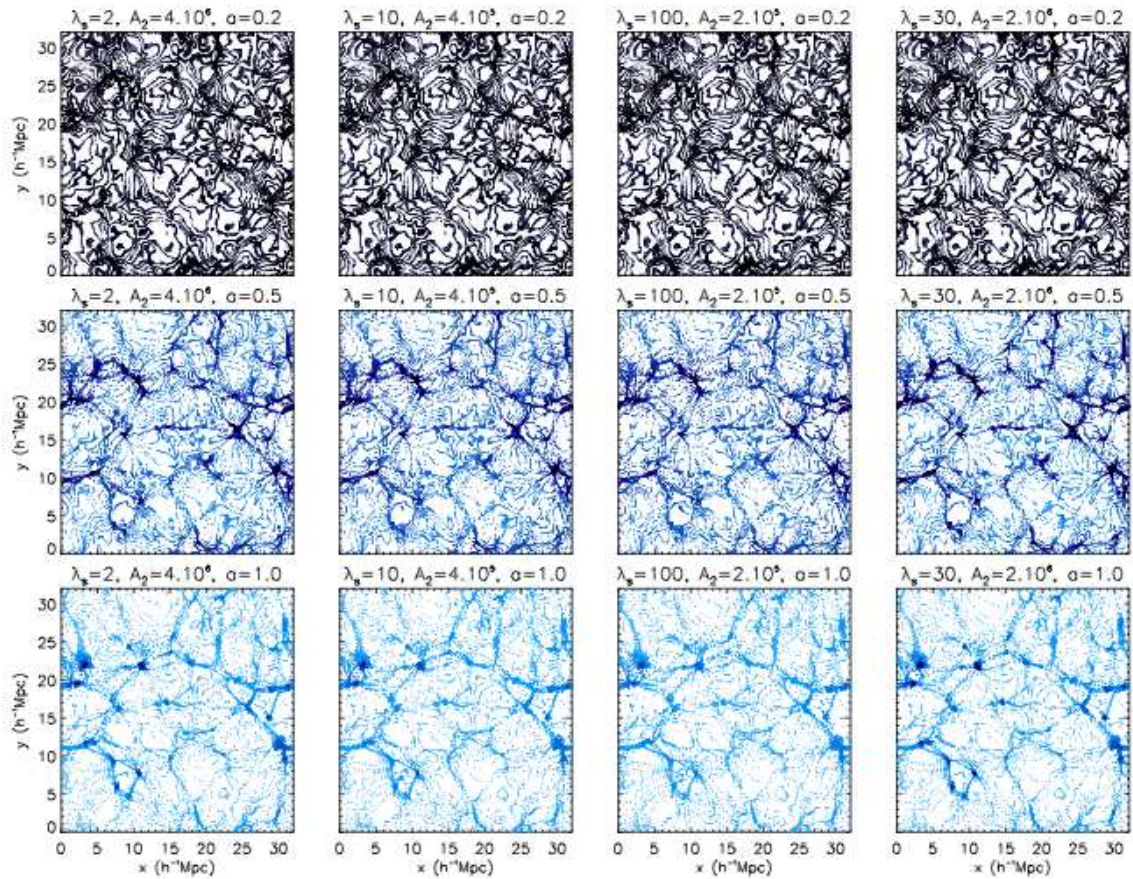}
\caption{(Colour Online) The colour scale plot of the value of $\beta$ as a function of coordinates $x, y$, for the same thin slice of the $32h^{-1}$~Mpc simulation box as in Fig.~\ref{fig:force}. We show this for the four models we have simulated and at three different output times, as given in the subtitle of each panel. Each point represents a particle, and the colour of the point depends on the value of $\beta$ at the position of that particle: for all the panels the lightest colour (white) denotes $\beta=0.3$ and the darkest colour (black) denotes $\beta=10^{-7}$; the blue colour is interpolated linearly between these two extremes}
\label{fig:beta}
\end{figure*}

\begin{figure*}
\includegraphics[scale=1.0]{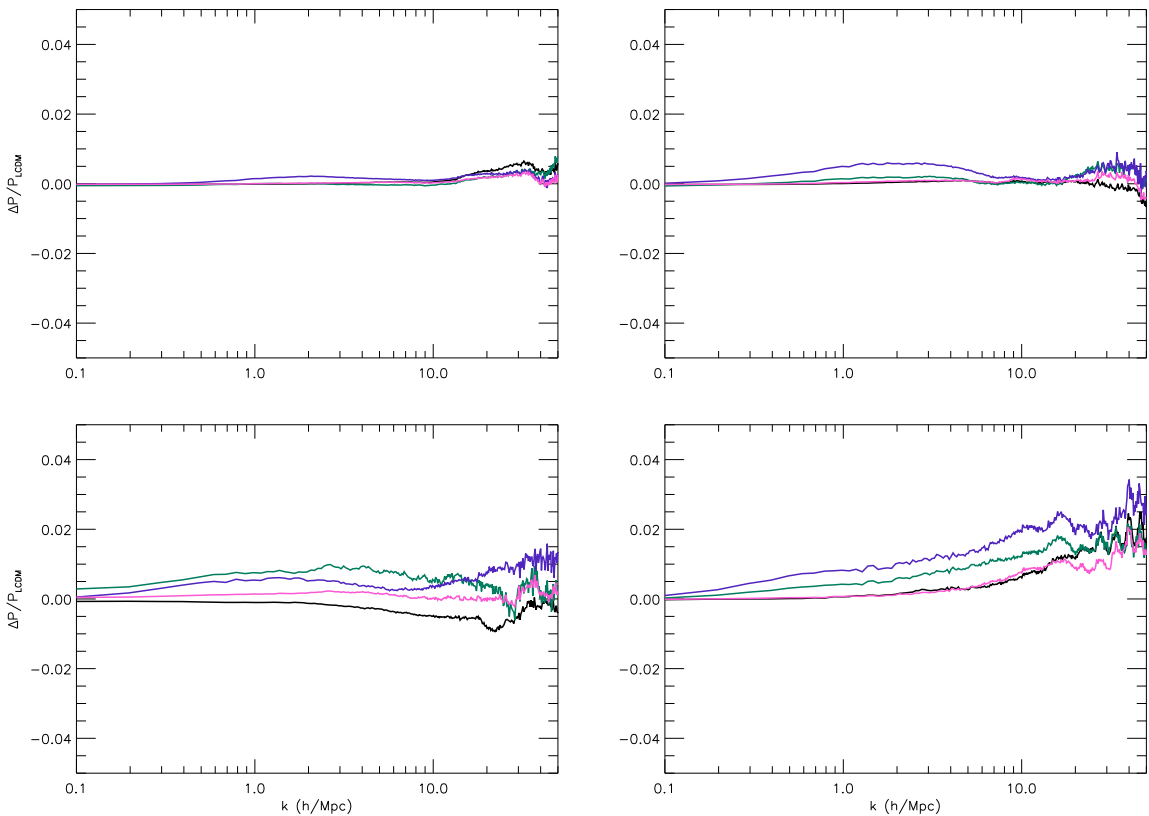}
\caption{(Colour Online) The fractional difference between the dilaton and $\Lambda$CDM nonlinear matter power spectra, which is defined to be $\left(P(k)-P_{\Lambda\rm{CDM}}(k)\right)/P_{\Lambda\rm{CDM}}(k)$. The black, green, pink and purple curves are respectively for the models with $(A_2,\lambda) = (4\times10^6,2)$, $(4\times10^5,10)$, $(2\times10^5,100)$ and $(2\times10^6,30)$. The four panels (upper-left, upper-right, lower-left and lower-right) are results at $a=0.3, 0.5, 0.7$ and $1.0$.}
\label{fig:power}
\end{figure*}

\begin{figure}
\includegraphics[scale=0.5]{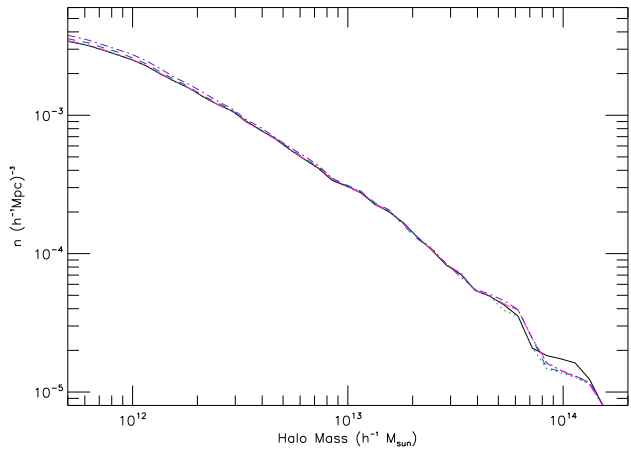}
\caption{(Colour Online) The mass functions for the $\Lambda$CDM model (black solid curve), and the four dilation models with $(A_2,\lambda) = (4\times10^6,2)$, $(4\times10^5,10)$, $(2\times10^5,100)$ and $(2\times10^6,30)$ (the coloured curves). All the curves are very close.}
\label{fig:mf}
\end{figure}

\begin{figure*}
\includegraphics[scale=1.0]{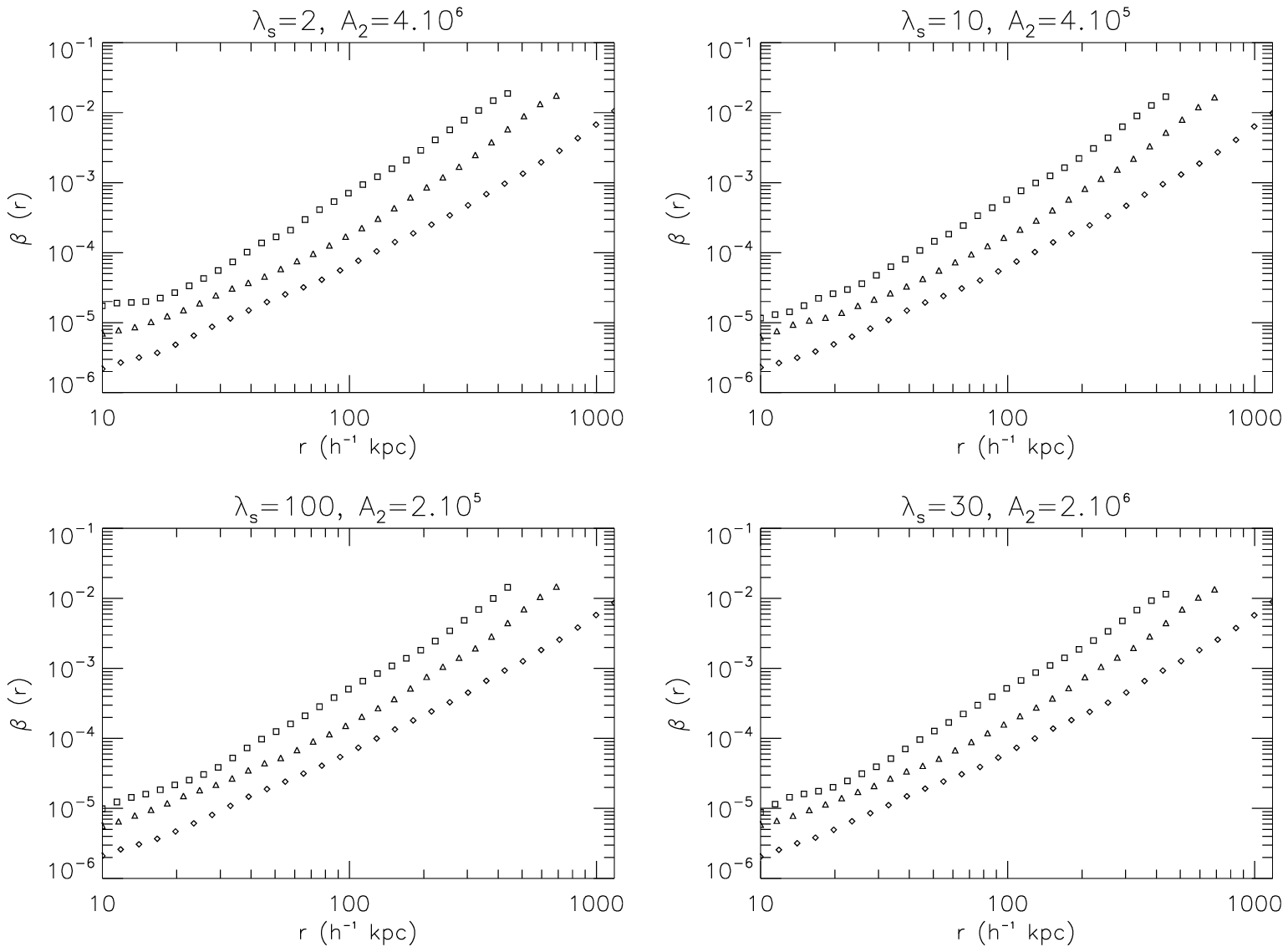}
\caption{The profiles for $\beta(r)$ in some chosen halos. The diamond, triangle and box represent results for three halos with masses equal to $3.5\times10^{14}$, $6.0\times10^{13}$ and $1.6\times10^{13}$ $h^{-1}M_{\bigodot}$ respectively. The horizontal axis is the distance from the halo centre, in unit of $h^{-1}$~kpc.}
\label{fig:profile}
\end{figure*}

\section{Numerical Results}

\label{sect:results}

In this section, we shall present our simulation results, including the snapshots, the matter power spectrum and the halo mass function.

\subsection{Snapshots}

As we have seen, in the dilaton model $\beta$ and thus the fifth force is suppressed in high density regions. In this subsection we demonstrate these qualitative features using some snapshots.

Fig.~\ref{fig:force} shows the comparison of the magnitudes of the fifth force and gravity for the four models we have considered, at three different output times $a=0.2, 0.5$ and $1.0$. For this, we pick out a thin slice from the middle of the $z=16h^{-1}$~Mpc simulation box, and compute the fifth force and gravity on the particles within that slice.

At early times, the density is high everywhere and we expect the fifth force on all particles in all the four models to be strongly suppressed, and this is confirmed by the first row of Fig.~\ref{fig:force}, which shows that the fifth force is much weaker than gravity. Note that the degree of suppression of the fifth force is dependent on the value of $A_2$: the larger $A_2$ is, the more the fifth force is suppressed.  Also, the fifth force is weaker in higher density regions (where gravity is stronger) than in lower density regions (where gravity is weaker), showing a strong dependence on the environment.

As the Universe expands, the overall density decreases and the fifth force becomes stronger, which could be seen in the lower rows of Fig.~\ref{fig:force}. If there is no suppression on the fifth force, then its strength should be
\ba\label{eq:fifth_to_grav}
\alpha &=& \frac{\bar\beta^2}{3\bar\beta^2+\lambda^{-2}}
\ea
times that of gravity \cite{Brax:2010}. For comparison, in the lowest row ($a=1$) we have over-plotted $\alpha$ times gravity as continuous curves. We can see that in the model with smaller $A_2$ (the middle two columns) Eq.~(\ref{eq:fifth_to_grav}) gives a relatively good description of the fifth force at least in some regions. But for the models with big $A_2$ (the first and fourth columns) the fifth force is strongly suppressed even today.

Since $\beta$ determines the strength of the fifth force [cf.~Eq.~\ref{eq:geodesic}], we are also interested in it. Fig.~\ref{fig:beta} shows the values of $\beta$ as a function of position in the same slices as Fig.~\ref{fig:force}. As expected, at very early times ($a=0.2$) $\beta\ll1$ because the fifth force is strongly suppressed. As the Universe expands, $\beta$ increases (the colour on the points becomes blue rather than black), but in the high density regions $\beta$ remains very small. Also, for the models with big $A_2$ (the first and fourth columns) the values of $\beta$ in high and low density regions tend to have stronger contrast, showing stronger environment-dependence. This is clearer in the third row, which shows the result at $a=1.0$. This is consistent with what we have seen in Fig.~\ref{fig:force}.

\subsection{Matter Power Spectrum}

The nonlinear matter power spectrum is an important structure formation observable and could be used to distinguish amongst different structure formation scenarios. In Ref.~\cite{Brax:2010} it has been shown that the growth rate of linear matter density perturbations in the dilaton model differs from that of $\Lambda$CDM only slightly, and therefore the dilaton model (with its parameters constrained by solar system tests) does not deviate at more than the percent level  from $\Lambda$CDM in practice. On the other hand, the fifth force in the dilaton model has a finite range and is expected to only take effect on the scales of galaxy clusters ($\sim\mathcal{O}$(Mpc)) and smaller, which already fall into the nonlinear regime. We are therefore interested in seeing how the fifth force affects the growth of density perturbations on these scales.

Fig.~\ref{fig:power} displays the fractional difference of the dilaton nonlinear matter power spectrum from that of the $\Lambda$CDM model, defined as $\left(P(k)-P_{\Lambda\rm{CDM}}(k)\right)/P_{\Lambda\rm{CDM}}(k)$. From this we can see that the difference is strongly suppressed even on small scales where the fifth force is expected to  take effect. This is different from the linear perturbation prediction of \cite{Brax:2010} (c.f.~Fig.~3 therein), which shows that the growth rate of density perturbation on small scales is significantly higher than that on large scales. The reason for this is that by linearising the scalar field equation, the nonlinearity of the dilaton model, which is the very mechanism that suppresses the fifth force in high density regions, is artificially removed (at least partially), and the strength of the fifth force is determined by the average, instead of the local, matter density. In contrast, the $N$-body simulation overcomes this problem by taking full account of the suppression of the fifth force.

The results indicate that it is even more difficult to use the nonlinear matter power spectrum to constrain the dilaton model or distinguish it from $\Lambda$CDM as the differences to the $\Lambda$CDM power spectrum are only a few percent on very small length scales at late times.

\subsection{Mass Function}

The halo mass function is another key structure formation observable. It is defined to be the number density of dark matter halos within a given mass range. Clearly, in case of a fifth force which could boost the clustering of matter, we expect more halos to form. In Fig.~\ref{fig:mf} we have shown the mass functions of the dilaton models compared with that of $\Lambda$CDM, at $z=0$. Although the dilaton models do have higher mass functions than $\Lambda$CDM, especially for small halos which generally live in low-density regions where the fifth force is less suppressed, the differences are again very small, making all these models hard to distinguish in practice at present and a challenge for future surveys.

\subsection{Halo Profile of $\beta$}
In Fig.~\ref{fig:profile} we show the profile of $\beta$ inside the dark matter halos, which are assumed to be spherical. Because $\beta$ characterises the strength of the fifth force, this can also provide information about the fifth force in halos. We have selected three halos with different masses (respectively $3.5\times10^{14}$, $6.0\times10^{13}$ and $1.6\times10^{13}$ solar mass) to check the results.

As can be seen from Fig.~\ref{fig:profile}, $\beta$ (and thus the strength of the fifth force) increases from inner to outer regions of the halos as the matter density is highest and the fifth force most severely suppressed in the central region. Furthermore, the fifth force is stronger for smaller halos, because those generally reside in low-density regions where the fifth force is less suppressed. However, for all these selected halos $\beta$ is at most $\sim\mathcal{O}(10^{-2})$ and typically less than $\sim\mathcal{O}(10^{-3})$ except near the halo edge, which mean that well inside the halos (such as where the solar system is) the fifth force is much weaker than gravity and has negligible effects (and we have not even included baryons in the simulations, which are generally much denser than dark matter in galaxies).

The results are consistent with what we have seen in the nonlinear matter power spectra and mass functions, all of them showing that the fifth force has little influence in the structure formation of the dilaton models (as long as solar system tests are passed).

\section{Summary and Conclusion}

\label{sect:con}

The dilaton model of \cite{Brax:2010} is an interesting alternative to the chameleon models with a different  mechanism by which the fifth force produced by a coupling between matter and scalar field(s) could be suppressed in high density regions. The theory therefore evades all solar system constraints while at the same time leaving open the possibility of significant effects on cosmological scales. The dilaton model has the advantage of being motivated from fundamental string theory. Given the parameter space allowed by local experiments, we have found that the effects of the fifth force on linear-perturbation evolution is weak. Here we have studied the possible imprints on the nonlinear evolution of large-scale structure using $N$-body simulations.

By solving the whole nonlinear equation instead of using linearisation, $N$-body simulations could fully capture the environment-dependence of the (scalar) dilaton field, and our results confirm the expectation that the high matter density in galaxy clusters strongly suppresses the strength of the fifth force. Consequently, the key cosmological observables such as the nonlinear matter power spectrum and mass function of the dilaton model are even closer to the corresponding $\Lambda$CDM predictions than that suggested by the linear perturbation analysis.

These results show that the  suppression of the environment-dependent coupling strength in the dilaton model is very efficient, and the model in practice satisfies all the known constraints, from solar system to cosmological. On the other hand, this also means that it is difficult to distinguish the dilaton model from the $\Lambda$CDM paradigm using the current (and possibly next generation of) cosmological observations. There may, however, be larger imprints of the fifth force in the galaxy clusters which reside in voids, where the overall density is low and the fifth force could be as strong as gravity. However, the spatial and mass resolutions of our simulations do not allow a detailed analysis of this.

Note that in the simulations of this work we have only included dark matter but not baryons. However, as long as the scalar field has a uniform coupling to different matter species, we expect that all the results will qualitatively remain. In particular, in the inner regions of the halos, baryon density is much higher than that of dark matter, which could further suppress the fifth force compared to what we have seen in our simulations.

In conclusion, while observationally hard to distinguish from the $\Lambda$CDM model, the environmentally dependent dilaton model is a very effective way to shield the dilaton from observations.

\acknowledgements

The $N$-body simulations in this work have been performed using a modified version of the publicly-available code {\tt MLAPM} \cite{Knebe:2001}, on the {\tt COSMOS} supercomputer of the UK. The nonlinear matter power spectrum is measured using the {\tt POWMES} code \cite{Colombi:2008dw}.  We thank Marco Baldi for comments. This work is supported in part by STFC. BL acknowledges the support of Queens' College, Cambridge.

\appendix

\section{Useful Expressions}

\label{appen:express}

Up to first order in the perturbed metric variables $\phi, \psi$, the nonzero components of the symmetric Levi-Civita connection are
\ba
\Gamma^0_{00} &=& \frac{a'}{a}+\phi',\\
\Gamma^0_{0\mu} &=& \phi_{,\mu},\\
\Gamma^\mu_{00} &=& \phi^{,\mu},\\
\Gamma^\mu_{0\nu} &=& \left(\frac{a'}{a}-\psi'\right)\delta^\mu_{\ \nu},\\
\Gamma^0_{\mu\nu} &=& \gamma_{\mu\nu}\left[\frac{a'}{a}(1-2\phi-2\psi)-\psi'\right],\\
\Gamma^\mu_{\nu\rho} &=& -\psi_{,\rho}\delta{\mu}_{\ \nu}-\psi_{,\nu}\delta^{\mu}_{\ \rho}+\psi^{,\mu}\gamma_{\nu\rho}.
\ea
The components of the Ricci tensor and Ricci scalar up to first order in $\phi,\psi$ are then easy to compute as
\ba
R_{00} &=& \phi^{,\mu}_{\ ,\mu}-3\left[\frac{a''}{a}-\left(\frac{a'}{a}\right)^2\right]+3\psi''\nonumber\\
&&+3\frac{a'}{a}\left(\phi'+\psi'\right),\\
R_{0\mu} &=& 2\psi'_{,\mu}+2\frac{a'}{a}\phi_{,\mu},\\
R_{\mu\nu} &=& -\psi''\gamma_{\mu\nu} - \frac{a'}{a}\left(\phi'+5\psi'\right)\gamma_{\mu\nu}-\psi^{,\rho}_{\ ,\rho}\gamma_{\mu\nu}\nonumber\\
&&+\left[\frac{a''}{a}+\left(\frac{a'}{a}\right)^2\right](1-2\phi-2\psi)\gamma_{\mu\nu}\nonumber\\
&&-(\phi-\psi)_{,\mu\nu},\\
R &=& \frac{6}{a^2}\frac{a''}{a}(1-2\phi) + \frac{1}{a^2}\left(4\psi^{,\mu}_{\ ,\mu}-2\phi^{,\mu}_{\ ,\mu}\right)\nonumber\\
&&-\frac{6}{a^2}\left[\psi''+\frac{a'}{a}\left(\phi'+3\psi'\right)\right].
\ea

\section{Discretisation of Equations}

\label{appen:discrete}

To implement the nonrelativistic equations into our numerical code, we have to rewrite them using code units, which are given by
\ba \mathbf{x}_c\ =\ \frac{\mathbf{x}}{B},\ \ \ \mathbf{p}_c\ =\ \frac{\mathbf{p}}{H_0{B}},\ \ \ t_c\ =\ tH_0,\nonumber\\
\Phi_c\ =\ \frac{\Phi}{(H_0{B})^{2}},\ \ \ \rho_c\ =\ \frac{\rho_m}{\bar{\rho}_m},\ \ \ \nabla_c\ =\ B\nabla_{\bf{x}}\ea
in which a subscript $_c$ denotes code unit, $B$ is the size of the simulation box and $H_0=100h$~km/s/Mpc. In what follows we shall write $\nabla=\nabla_c$ for simplicity.

\subsection{Scalar Field Equation of Motion}

Recall that in our model we have chosen $\varphi_0=0$ such that $\beta(\varphi)=A_2\varphi$. It is then the same to solve for $\varphi$ or to solve for $\beta$. As $A_2\gg1$, $\beta\gg\varphi$ and so we shall solve for $\beta$ rather than $\varphi$ to reduce possible numerical errors. The equation of motion for $\beta$ could be obtained simply by multiplying that for $\varphi$ by $A_2$:
\ba\label{eq:sfeom_beta}
&&\nabla_{\bf{x}}\cdot\left[k(\beta)\nabla_{\bf{x}}\beta\right]\nonumber\\
&\approx& \frac{4\pi GA_2a^2}{k(\beta)}\left[\beta\left\{A(\beta)\rho_{\rm{m}}+4V(\beta)\right]-V(\beta)\right\}\nonumber\\
&&-\frac{4\pi GA_2a^2}{k(\bar{\beta})}\left[\bar{\beta}\left\{A(\bar{\beta})\bar{\rho}_{\rm{m}}+4V(\bar{\beta})\right]-V(\bar{\beta})\right\},
\ea
where we have used $\beta$ instead of $\varphi$ as the variable.

As discussed in \cite{Brax:2010}, $\beta$ characterises the strength of the fifth force. In high density environments, $\beta\ll1$ so that the fifth force is too weak to be measured; in low density regions, however, we have $\beta\sim0.23$ today, indicating that the fifth force is roughly as strong as gravity. Furthermore, Eq.~(\ref{eq:sfeom_beta}) does not say anything about the sign of $\beta$.

Obviously, because $\beta$ ranges from $\mathcal{O}\left(10^{-6}\right)$ to $\mathcal{O}(1)$, using $\beta$ directly in the numerical code could easily cause big numerical errors in the regions where $\beta$ is small. One alternative is to use  $\ln(\beta)$ as a new variable, but this does not necessarily work because $\beta$ might be negative. Therefore, in this work, we shall use a different variable $u\equiv\beta^{1/n}$, with $n$ being some odd positive integer, as the redefined scalar field. More explicitly, we shall adopt $n=9$ which guarantees that $u\sim\mathcal{O}(0.1-1)$, \ie, $u$ spans a much smaller range than $\beta$, making it easier to control numerical errors. Furthermore, $n$ being odd makes sure that $u$ is never undefined even if $\beta<0$.

In terms of $u$, we have
\ba
k(u) &=& \sqrt{3u^{2n}+\lambda^{-2}},\\
A(u) &=& 1+\frac{u^{2n}}{2A_2},\\
V(u) &=& \left(1+\frac{u^{2n}}{2A_2}\right)^4V_0\exp\left(-\frac{u^n}{A_2}\right).
\ea
Then, defining 
\ba
\tilde{\lambda} &\equiv& \frac{8\pi GV_0}{3H_0^2},
\ea
and using the code units defined above, we could rewrite the scalar equation of motion Eq.~(\ref{eq:sfeom_beta}) as
\begin{widetext}
\ba\label{eq:sfeom_u}
&&\frac{ac^2}{\left(BH_0\right)^2}\nabla\cdot(b\nabla u)\nonumber\\
&\approx& \frac{A_2}{k(u)}\left\{\left[\frac{3}{2}\left(1+\frac{u^{2n}}{2A_2}\right)\Omega_{\rm{m}}\rho_c+6\tilde{\lambda} a^3\left(1+\frac{u^{2n}}{2A_2}\right)^4\exp\left(-\frac{u^n}{A_2}\right)\right]u^n
-\frac{3}{2}\tilde{\lambda} a^3\left(1+\frac{u^{2n}}{2A_2}\right)^4\exp\left(-\frac{u^n}{A_2}\right)\right\}\nonumber\\
&&-\frac{A_2}{k(\bar{\beta})}\left\{\left[\frac{3}{2}\left(1+\frac{\bar{\beta}^2}{2A_2}\right)\Omega_{\rm{m}}+6\tilde{\lambda} a^3\left(1+\frac{\bar{\beta}^2}{2A_2}\right)^4\exp\left(-\frac{\bar{\beta}}{A_2}\right)\right]\bar{\beta}-\frac{3}{2}\tilde{\lambda} a^3\left(1+\frac{\bar{\beta}^2}{2A_2}\right)^4\exp\left(-\frac{\bar{\beta}}{A_2}\right)\right\}
\ea
\end{widetext}
where we have defined
\ba
b(u) &=& nu^{n-1}\sqrt{3u^{2n}+\lambda^{-2}},
\ea
and $\bar{\beta}$ is the background value of $\beta$, which can be computed as \cite{Brax:2010}
\ba
\bar{\beta} &=& \frac{\Omega_\Lambda a^3}{\Omega_{\rm{m}}+4\Omega_\Lambda a^3}.
\ea

The full equation for $u$, Eq.~(\ref{eq:sfeom_u}), contains the quantity $\nabla\cdot\left(b\nabla u\right)$. To discretise it, we shall assume that the discretisation is performed on a grid with grid spacing $h$. We shall require second order precision which is the same as the default Poisson solver in {\tt MLAPM}, and then $\nabla u$ in one dimension can be written as
\be
\nabla u\ \rightarrow\ \nabla^h u_j\ =\ \frac{u_{j+1}-u_{j-1}}{2h}
\ee
where a subscript $j$ means that the quantity is evaluated on the $j$-th point. The generalisation to the three dimensional case is straightforward.

The factor $b$ in $\nabla\cdot(b\nabla u)$ makes this a standard variable coefficient problem. We need also to discretise $b$, and do it in this way (again for one dimension) \cite{Li:2009sy}:
\begin{eqnarray}
&&\nabla\cdot(b\nabla u)\nonumber\\
&\rightarrow& \frac{1}{h^2}\left[b_{j+\frac{1}{2}}u_{j+1} - u_j\left(b_{j+\frac{1}{2}}+b_{j-\frac{1}{2}}\right) + b_{j-\frac{1}{2}}u_{j-1}\right], \ \ \
\end{eqnarray}
in which $b_{j\pm\frac{1}{2}}=\frac{1}{2}\left(b_j+b_{j\pm1}\right)$. Generalising this to three dimensions, we have
\begin{widetext}
\begin{eqnarray}
\nabla\cdot(b\nabla u)
&\rightarrow& \frac{1}{h^2}\left[b_{i+\frac{1}{2},j,k}u_{i+1,j,k} - u_{i,j,k}\left(b_{i+\frac{1}{2},j,k}+b_{i-\frac{1}{2},j,k}\right) + b_{i-\frac{1}{2},j,k}u_{i-1,j,k}\right]\nonumber\\
&&+\frac{1}{h^2}\left[b_{i,j+\frac{1}{2},k}u_{i,j+1,k} - u_{i,j,k}\left(b_{i,j+\frac{1}{2},k}+b_{i,j-\frac{1}{2},k}\right) + b_{i,j-\frac{1}{2},k}u_{i,j-1,k}\right]\nonumber\\
&&+\frac{1}{h^2}\left[b_{i,j,k+\frac{1}{2}}u_{i,j,k+1} - u_{i,j,k}\left(b_{i,j,k+\frac{1}{2}}+b_{i,j,k-\frac{1}{2}}\right) + b_{i,j,k-\frac{1}{2}}u_{i,j,k-1}\right].
\end{eqnarray}
\end{widetext}
Then the discrete version of Eq.~(\ref{eq:sfeom_u}) is
\begin{eqnarray}\label{eq:diffop}
L^{h}\left(u_{i,j,k}\right) &=& 0,
\end{eqnarray}
in which
\begin{widetext}
\ba\label{eq:diffop1}
L^{h}\left(u_{i,j,k}\right) &=&
\frac{1}{h^{2}}\frac{ac^2}{(BH_0)^2}\left[b_{i+\frac{1}{2},j,k}u_{i+1,j,k}
- u_{i,j,k}\left(b_{i+\frac{1}{2},j,k}+b_{i-\frac{1}{2},j,k}\right)
+ b_{i-\frac{1}{2},j,k}u_{i-1,j,k}\right]\nonumber\\
&& +
\frac{1}{h^{2}}\frac{ac^2}{(BH_0)^2}\left[b_{i,j+\frac{1}{2},k}u_{i,j+1,k}
- u_{i,j,k}\left(b_{i,j+\frac{1}{2},k}+b_{i,j-\frac{1}{2},k}\right)
+ b_{i,j-\frac{1}{2},k}u_{i,j-1,k}\right]\nonumber\\
&& +
\frac{1}{h^{2}}\frac{ac^2}{(BH_0)^2}\left[b_{i,j,k+\frac{1}{2}}u_{i,j,k+1}
- u_{i,j,k}\left(b_{i,j,k+\frac{1}{2}}+b_{i,j,k-\frac{1}{2}}\right)
+ b_{i,j,k-\frac{1}{2}}u_{i,j,k-1}\right]\nonumber\\
&&-\frac{A_2}{\sqrt{3u_{i,j,k}^{2n}+\lambda^{-2}}}\left[\frac{3}{2}\left(1+\frac{u_{i,j,k}^{2n}}{2A_2}\right)\Omega_{\rm{m}}\rho_c+6\tilde{\lambda} a^3\left(1+\frac{u_{i,j,k}^{2n}}{2A_2}\right)^4\exp\left(-\frac{u_{i,j,k}^n}{A_2}\right)\right]u_{i,j,k}^n\nonumber\\
&&+\frac{A_2}{\sqrt{3u_{i,j,k}^{2n}+\lambda^{-2}}}\frac{3}{2}\tilde{\lambda} a^3\left(1+\frac{u_{i,j,k}^{2n}}{2A_2}\right)^4\exp\left(-\frac{u_{i,j,k}^n}{A_2}\right)\nonumber\\
&&+\frac{A_2}{\sqrt{3\bar{\beta}^2+\lambda^{-2}}}\left[\frac{3}{2}\left(1+\frac{\bar{\beta}^2}{2A_2}\right)\Omega_{\rm{m}}+6\tilde{\lambda} a^3\left(1+\frac{\bar{\beta}^2}{2A_2}\right)^4\exp\left(-\frac{\bar{\beta}}{A_2}\right)\right]\bar{\beta}\nonumber\\
&&-\frac{A_2}{\sqrt{3\bar{\beta}^2+\lambda^{-2}}}\frac{3}{2}\tilde{\lambda} a^3\left(1+\frac{\bar{\beta}^2}{2A_2}\right)^4\exp\left(-\frac{\bar{\beta}}{A_2}\right).
\ea
\end{widetext}
Then the Newton-Gauss-Seidel iteration says that we can obtain a new (and usually more accurate) solution of $u$, $u^{\rm new}_{i,j,k}$, using our knowledge about the old (and less acurate) solution $u^{\rm old}_{i,j,k}$ as
\begin{eqnarray}\label{eq:GS}
u^{\mathrm{new}}_{i,j,k} &=& u^{\mathrm{old}}_{i,j,k} -
\frac{L^{h}\left(u^{\mathrm{old}}_{i,j,k}\right)}{\partial
L^{h}\left(u^{\mathrm{old}}_{i,j,k}\right)/\partial u_{i,j,k}}.
\end{eqnarray}
The old solution will be replaced with the new one once the latter is ready, using a red-black Gauss-Seidel sweeping scheme. Note that
\begin{widetext}
\begin{eqnarray}\label{eq:diffop2}
\frac{\partial L^{h}(u_{i,j,k})}{\partial u_{i,j,k}} &=&
\frac{1}{2h^{2}}d\left(u_{i,j,k}\right)\frac{ac^2}{(BH_0)^2}\left[u_{i+1,j,k}+u_{i-1,j,k}+u_{i,j+1,k}
+u_{i,j-1,k}+u_{i,j,k+1}+u_{i,j,k-1}-6u_{i,j,k}\right]\nonumber\\
&&-\frac{1}{2h^{2}}\frac{ac^2}{(BH_0)^2}\left[b_{i+1,j,k}+b_{i-1,j,k}+b_{i,j+1,k}
+b_{i,j-1,k}+b_{i,j,k+1}+b_{i,j,k-1}+6b_{i,j,k}\right]\nonumber\\
&&-\frac{A_2nu_{i,j,k}^{n-1}}{\sqrt{3u^{2n}_{i,j,k}+\lambda^{-2}}}\left[\frac{3}{2}\left(1+\frac{u^{2n}_{i,j,k}}{2A_2}\right)\Omega_{\rm{m}}\rho_c+6\tilde{\lambda} a^3\left(1+\frac{u^{2n}_{i,j,k}}{2A_2}\right)^4\exp\left(-\frac{u^n_{i,j,k}}{A_2}\right)\right]\nonumber\\
&&-\frac{\frac{3}{2}\tilde{\lambda} a^3nu_{i,j,k}^{n-1}}{\sqrt{3u^{2n}_{i,j,k}+\lambda^{-2}}}\left[\left(1+\frac{u^{2n}_{i,j,k}}{2A_2}\right)+4u_{i,j,k}^{n}\right]\left(1+\frac{u^{2n}_{i,j,k}}{2A_2}\right)^3\exp\left(-\frac{u^n_{i,j,k}}{A_2}\right)\nonumber\\
&&-\frac{nu^{2n-1}_{i,j,k}}{\sqrt{3u^{2n}_{i,j,k}+\lambda^{-2}}}\left[\frac{3}{2}\Omega_{\rm{m}}\rho_cu^{n}_{i,j,k}-6\tilde{\lambda} a^3\left(1+\frac{u^{2n}_{i,j,k}}{2A_2}\right)^4\exp\left(-\frac{u^n_{i,j,k}}{A_2}\right)\right]\nonumber\\
&&+\frac{24\tilde{\lambda} a^3nu^{3n-1}_{i,j,k}}{\sqrt{3u^{2n}_{i,j,k}+\lambda^{-2}}}\left(1+\frac{u^{2n}_{i,j,k}}{2A_2}\right)^3\exp\left(-\frac{u^n_{i,j,k}}{A_2}\right)\nonumber\\
&&+\frac{3nA_2u^{2n-1}_{i,j,k}}{\left(3u_{i,j,k}^{2n}+\lambda^{-2}\right)^{3/2}}\left[\frac{3}{2}\left(1+\frac{u_{i,j,k}^{2n}}{2A_2}\right)\Omega_{\rm{m}}\rho_c+6\tilde{\lambda} a^3\left(1+\frac{u_{i,j,k}^{2n}}{2A_2}\right)^4\exp\left(-\frac{u_{i,j,k}^n}{A_2}\right)\right]u_{i,j,k}^n\nonumber\\
&&-\frac{3nA_2u^{2n-1}_{i,j,k}}{\left(3u_{i,j,k}^{2n}+\lambda^{-2}\right)^{3/2}}\frac{3}{2}\tilde{\lambda} a^3\left(1+\frac{u_{i,j,k}^{2n}}{2A_2}\right)^4\exp\left(-\frac{u_{i,j,k}^n}{A_2}\right)
\end{eqnarray}
\end{widetext}
where we have defined
\ba
d(u) &\equiv& \frac{db(u)}{du}\nonumber\\
&=& \frac{3n(2n-1)u^{3n-2}+n(n-1)\lambda^{-2}u^{n-2}}{\sqrt{3u^{2n}+\lambda^{-2}}}.
\ea

In principle, if we start from some high redshift, then the initial guess of $u_{i,j,k}$ could be chosen as the background value because we expect that any perturbations should be small then. For subsequent time steps we can use either the solution at the last time step or some analytical approximated solution as the initial guess.

\subsection{Poisson Equation}

In terms of the newly-defined scalar field $u$ and using the code units, the modified Poisson equation becomes
\begin{widetext}
\ba\label{eq:poisson_dis}
\nabla^2\Phi_c &=& \frac{3}{2}\Omega_{\rm{m}}\left[\left(1+\frac{u_{i,j,k}^{2n}}{2A_2}\right)\rho_{c,i,j,k}-\left(1+\frac{\bar{\beta}^{2}}{2A_2}\right)\right]\nonumber\\
&&-3\tilde{\lambda} a^3\left[\left(1+\frac{u_{i,j,k}^{2n}}{2A_2}\right)^4\exp\left(-\frac{u^n_{i,j,k}}{A_2}\right)-\left(1+\frac{\bar{\beta}^{2}}{2A_2}\right)^4\exp\left(-\frac{\bar{\beta}}{A_2}\right)\right].
\ea
\end{widetext}

The discretisation of $\nabla^2\Phi_c$ is straightforward and will not be presented here.

\subsection{Particle Equation of Motion}

Using the code units, Eq.~(\ref{eq:dxdt}) could be easily rewritten as
\ba\label{eq:dxdt_dis}
\frac{d\bf{x}_c}{dt_c} &=& \frac{\bf{p}_c}{a^2}.
\ea
Similarly, Eq.~(\ref{eq:dpdt}) becomes
\ba\label{eq:dpdt_dis}
\frac{d\bf{p}_c}{dt_c} &=& -\frac{1}{a}\nabla\Phi_c - \frac{1}{a}\frac{nu^{2n-1}_{i,j,k}}{A_2}\frac{ac^2}{\left(BH_0\right)^2}\nabla u\nonumber\\
&&-\frac{1}{A_2}u^n_{i,j,k}\frac{a\dot{\beta}}{H_0}\bf{p}_c.
\ea


\begin{thebibliography}{99}

\bibitem{rubakov}  
V.~A.~ Rubakov and  P.~G.~Tinyakov, Phys.~Usp.~{\bf51} 759 (2008) and references therein. 

\bibitem{woodard}
R.~P.~Woodard, Lect.~Notes~Phys.~{\bf720}, 403 (2007).
 

\bibitem{Tolley}
C.~de Rham, G.~Gabadadze and A.~J.~Tolley,
arXiv:1011.1232 [hep-th].

\bibitem{DGP}
  G.~Dvali, G.~Gabadadze and M.~Porrati, Phys.~Lett.~B{\bf 485}, 208 (2000).

\bibitem{fR}
  T.~P.~Sotiriou and V.~Faraoni, Rev.~Mod.~Phys.~{\bf 82}, 451 (2010).

\bibitem{scalar_tensor}
  {\it The Scalar-Tensor Theory of Gravitation}, Y.~Fujii and K.~-i~Maeda, Cambridge University Press, 2003.

\bibitem{Khoury:2004a}
  J.~Khoury and A.~Weltman, Phys.~Rev.~Lett.~{\bf 93}, 171104 (2004).

\bibitem{Khoury:2004}
  J.~Khoury and A.~Weltman, Phys.~Rev.~D{\bf 69}, 044026 (2004).

\bibitem{Mota:2006}
  D.~F.~Mota and D.~G.~Shaw, Phys.~Rev.~Lett.~{\bf 97}, 151102 (2006).

\bibitem{Mota:2007}
  D.~F.~Mota and D.~G.~Shaw, Phys.~Rev.~D{\bf 75}, 063501 (2007).

\bibitem{Li:2007}
  B.~Li and J.~D.~Barrow, Phys.~Rev.~D{\bf 75}, 084010 (2007).

\bibitem{Brax:2008}
  P.~Brax, C.~Van~de~Bruck, A.-C.~Davis and D.~J.~Shaw Phys.~Rev.~D{\bf 78}, 104021 (2008).

\bibitem{Carroll}
  S.~M.~Carroll,
  Phys.\ Rev.\ Lett.\  {\bf 81} (1998) 3067
  [arXiv:astro-ph/9806099].

\bibitem{Chow:2009}
  N.~Chow and J.~Khoury, Phys.~Rev.~D{\bf 80}, 024037 (2009).

\bibitem{DeFelice:2010}
  A.~De~Felice and S.~Tsujikawa, Phys.~Rev.~Lett.~{\bf 105}, 111301 (2010).

\bibitem{Vainshtein:1972}
  A.~I.~Vainshtein, Phys.~Lett.~B{\bf 39}, 393 (1972).

\bibitem{dilaton1}
  M.~Gasperini, F.~Piazza and G.~Veneziano,
  Phys.\ Rev.\  D {\bf 65} (2002) 023508
  [arXiv:gr-qc/0108016].

\bibitem{dilaton2}
  T.~Damour, F.~Piazza and G.~Veneziano,
  Phys.\ Rev.\ Lett.\  {\bf 89} (2002) 081601
  [arXiv:gr-qc/0204094].

\bibitem{dilaton3}
  T.~Damour, F.~Piazza and G.~Veneziano,
  Phys.\ Rev.\  D {\bf 66} (2002) 046007
  [arXiv:hep-th/0205111].


\bibitem{Damour:1994}
  T.~Damour and A.~M.~Polyakov, Nucl.~Phys.~B{\bf 423}, 532 (1994).

\bibitem{Brax:2010}
  P.~Brax, C.~van~de~Bruck, A.-C.~Davis and D.~J.~Shaw (2010), arXiv:1005.3735 [astro-ph.CO].
  
\bibitem{Brax:2011}
  F.~Bernardeau and P.~Brax (2011), arXiv:1102.1907 [astro-ph.CO].

\bibitem{Li:2009sy}
  B.~Li and H.~Zhao, Phys.\ Rev.\  D {\bf 80}, 044027 (2009).

\bibitem{Li:2010mq}
  B.~Li and H.~Zhao, Phys.\ Rev.\  D {\bf 81}, 104047 (2009).

\bibitem{Li:2010nc}
  B.~Li and J.~D.~Barrow, Phys.\ Rev.\ D  {\bf 83}, 024007 (2011).

\bibitem{Li:2010st}
  B.~Li, D.~F.~Mota and J.~D.~Barrow, Astrophys.~J., {\bf 728}, 109 (2011).

\bibitem{Knebe:2001}
  A.~Knebe, A.~Green and J.~Binney,
  Mon.~Not.~R.~Astron.~Soc., {\bf 325}, 845 (2001).

\bibitem{Press:1988}
  W.~H.~Press, B.~P.~Flannery, S.~A.~Teukolsky and W.~T.~Vetterling,
  {\it Numerical Recipes in C: The Art of Scientific Computing}, Cambridge University Press, 1988.

\bibitem{Oyaizu:2008}
  H.~Oyaizu,
  Phys.\ Rev.\  D {\bf 78}, 123523 (2008).

\bibitem{Colombi:2008dw}
  S.~Colombi, A.~H.~Jaffe, D.~Novikov and C.~Pichon, Mon.~Not.~R.~Astron.~Soc.~{\bf 303}, 511 (2009).

\bibitem{Bertschinger:1995er}
  E.~Bertschinger, arXiv:astro-ph/9506070.

\bibitem{Gill:2004km}
  S.~P.~D.~Gill, A.~Knebe and B.~K.~Gibson,
  Mon.\ Not.\ Roy.\ Astron.\ Soc.\  {\bf 351}, 399 (2004).

\end{thebibliography}
\end{document}